\documentstyle{mn} 
\input epsf
\title[X-ray wakes]
{X-ray wakes as probes of galaxy cluster dynamics}
\author[M. R. Merrifield]
          {Michael. R.~Merrifield \\
          Department of Physics and Astronomy\\
          University of Southampton,\\
          Southampton SO17 1BJ}
\begin{document}
\maketitle

\begin{abstract} 

If a galaxy resides in a cluster, then its passage through the
pervasive intracluster medium will produce a detectable signature in
the X-ray emission from the cluster.  Such features have now been
detected in a number of systems.  The simplest kinematic information
that can be extracted from this signature is the galaxy's direction of
motion on the plane of the sky.  This paper explores the constraints
on cluster dynamics that could be derived from such information.  In
particular, we show that it is possible to define a projected
anisotropy parameter, $B(R)$, which is directly analogous to the usual
orbital anisotropy parameter.  We describe an estimator for this
quantity, $\hat B(R)$, which can be derived in a robust and
straightforward manner.  We present a simple dynamical model for a
cluster consisting of a Michie distribution function of galaxies
orbiting in a truncated singular isothermal sphere potential.  Using
this model, we demonstrate the ambiguity between the distribution of
mass and the distribution of galaxy orbits when interpreting the
traditional measures of cluster kinematics (the projected density of
galaxies and their line-of-sight velocity dispersion).  As an example,
we show how two very different dynamical models can fit the kinematic
properties of the Coma cluster.  We demonstrate that the measurement
of $\hat B$ using a relatively small sample of wake directions
($N_{\rm wake} \approx 50$) would provide an effective mechanism for
lifting this degeneracy.  Thus, by combining X-ray measurements of
wake directions with number counts and line-of-sight velocities
derived from optical data, it will prove possible to measure both the
orbit distribution and the form of the gravitational potential in
clusters of galaxies.  The requisite X-ray observations lie within
reach of the soon-to-be-launched AXAF satellite.

\end{abstract}


\section{Introduction}

As the seminal work on the Coma cluster by Kent \& Gunn (1982) first
demonstrated, it is possible to study the dynamics of a cluster of
galaxies by combining photometric observations of the galaxies'
positions on the plane of the sky with spectroscopic measurements that
reveal the line-of-sight velocities of cluster members.  If it is
assumed that the cluster is spherically symmetric, these observed
properties can usefully be quantified by the projected number density
of galaxies as a function of projected radius, $N(R)$, and the RMS
dispersion in the galaxies' line-of-sight velocities as a function of
projected radius, $\sigma_{\rm los}(R)$.  Dynamical models with
different distributions of galaxy orbits can then be tested against
the observations by comparing $N(R)$ and $\sigma_{\rm los}(R)$ to the
forms for these functions that the models predict.

In principle, such a study allows us to investigate the manner in
which a cluster has formed and evolved.  Galaxies have relatively long
dynamical ``memories,'' and so their current orbits must to some
degree reflect the processes by which the cluster formed.  For
example, if the galaxies are found to follow preferentially radial
orbits, then their motions presumably reflect the radial infall from
which the cluster originally formed.  Further, the kinematics of the
galaxies are dictated by the gravitational potential of the cluster as
a function of radius, $\Phi(r)$, and so dynamical modelling allows us
to probe the distribution of mass in the cluster.  Since the mass of a
galaxy cluster is dominated by dark matter, no study of the evolution
of such a system is complete without addressing the distribution of
this massive component as inferred from the gravitational potential.

Unfortunately, there is a fundamental ambiguity in such studies.  As
Binney \& Mamon (1982) demonstrated, for any given $\Phi(r)$ it is
possible to reproduce a wide range of functions $N(R)$ and
$\sigma_{\rm los}(R)$ simply by varying the balance between radial and
tangential motions in the orbits that the galaxies follow.  They
parameterized this balance by the anisotropy parameter,
\begin{equation}
\beta(r) = 1 - {\sigma_t^2 \over \sigma_r^2},
\label{eqn:beta}
\end{equation}
where $\sigma_t$ and $\sigma_r$ are the RMS velocities in the
tangential and radial directions for galaxies at radius $r$ in the
cluster.\footnote{Note that there are two equal tangential components
in a spherical system (in the $\theta$ and $\phi$ directions of a
spherical polar coordinate system), and so $\sigma_t^2$ is sometimes
defined to be a factor of two larger.}  Thus, it is not possible to
use just $N(R)$ and $\sigma_{\rm los}(R)$ to solve for both $\Phi(r)$
and $\beta(r)$.

One approach to lifting this degeneracy is to find some independent
technique for measuring $\Phi(r)$, and then use $N(R)$ and
$\sigma_{\rm los}(R)$ to solve for the orbital anisotropy, $\beta(r)$.
For example, the form of the gravitational potential can be inferred
from the properties of the X-ray emitting intracluster medium (ICM).
By measuring the temperature of this hot gas and its spatial
distribution using an X-ray telescope, it is possible to solve
uniquely for the form of the gravitational potential that contains the
gas -- see, for example, Eyles et al.\ (1991).  Alternatively, one can
use the distortion of optical images of background galaxies due to the
gravitational lensing effects of the cluster mass to measure $\Phi(r)$
directly (Tyson, Wenk \& Valdes 1990).

In practice, neither of these techniques is entirely satisfactory.
The measurement of $\Phi(r)$ using X-ray data requires one to measure
the temperature profile of the ICM as well as its density profile, and
the poor spatial and spectral resolution and limited sensitivity of
existing X-ray observations make this a difficult task.  The inversion
from gravitational lens distortions to $\Phi(r)$ can only be made if
the distorted shapes of many background galaxies are measured, and is
to some extent dependent on the assumed redshift distribution of the
background galaxies.  Further, neither of these techniques is
well-suited to measuring $\Phi(R)$ in poorer clusters, where the X-ray
emission will be weak, and the amount of gravitational-lens distortion
will be small.

Perhaps this problem can be more profitably attacked from the other
direction: if we can find some additional constraint on the
distribution of orbits in a cluster, then we can use this information
to lift the ambiguity and solve for both the galaxy kinematics and the
cluster potential.  One attempt to adopt this approach was made by
Merritt (1987).  In a further study of the Coma cluster's kinematics,
he showed that the distribution of member galaxies' line-of-sight
velocities will have a shape that may depart dramatically from a
Gaussian depending on the distribution of galaxy orbits.
Specifically, if the orbits are preferentially radial then the the
velocity distribution will have longer tails than a Gaussian, while if
the orbits are more circular it will have a ``boxier'' appearance.
This information in the shape of the line-of-sight velocity
distribution has been successfully exploited in dynamical studies of
the stars in individual galaxies (e.g. van der Marel \& Franx 1993,
Kuijken \& Merrifield 1993), but it has proved less useful in the case
of clusters.  The main difficulty in using the shape of a cluster's
velocity distribution is that clusters are not entirely relaxed
systems.  A relatively small amount of substructure in the
distribution of galaxy velocities within a cluster, perhaps due to an
infalling group of galaxies that has yet to disrupt completely, can
totally dominate the departures from a Gaussian in the line-of-sight
velocity distribution (Zabludoff, Franx \& Geller 1993).

In this paper, we explore an alternative candidate for the requisite
kinematic constraint: the X-ray signature of the interactions between
cluster galaxies and the surrounding ICM.  As a galaxy moves through
the ICM, the combination of gravitational focusing of the ICM and ram
pressure stripping of the galaxy's own interstellar medium (ISM) will
produce an enhancement in the density of gas behind the galaxy -- see
Balsara, Livio \& O'Dea (1994) for a review of these processes and the
results of some simulations.  Thus, we expect a galaxy's direction of
motion on the plane of the sky to be marked out in X-ray observations
by a ``wake'' of excess emission trailing behind it.  This phenomenon
has now been observed in several systems.  Sakelliou, Merrifield \&
McHardy (1997) found a wake of X-ray emission emanating from the radio
galaxy 4C34.16.  In this case, the bent morphology of the radio jets
provides a further measure of the galaxy's direction of travel, and
the wake is found to lie downstream from the moving galaxy.  A further
example is provided by the dumb-bell galaxy NGC~4782/3 (Colina \&
Borne 1995).  In this interacting pair of elliptical galaxies, the
X-ray emission from the individual components is offset from the
optical light produced by the galaxies in the sense that one
would expect if it were trailing behind the orbiting galaxies.  The
X-ray emission also reveals more extended wake features lagging behind
the orbiting components.  Finally, Jones et al.\ (1997) have found
that the X-ray emission from the elliptical galaxy NGC~1404 is
strongly distorted, forming a wake pointing away from the centre of
the Fornax cluster in which it resides.  They suggest that this X-ray
morphology arises from the ram pressure stripping of the galaxy's ISM,
and hence conclude that NGC~1404 is on a plunging radial orbit.

With the advent of more sensitive X-ray telescopes such as XMM and
high resolution imaging instruments such as AXAF, the detection of
these wake phenomena can only become more commonplace, and we might
reasonably expect to be able to use wakes as a probe of the kinematics
of many galaxies within a single cluster.  For example, Balsara et
al.\ (1994) have calculated that at the distance of the Coma cluster
one might expect to see many wake structures with extents of $\sim 2$
arcseconds -- a scale readily resolvable with AXAF.

In principle, a great deal of information can be gleaned from the
pattern of shocks and other structure present in X-ray wakes.
However, in addition to the velocity of the galaxy, this structure
depends in a complex manner on the properties of the ICM, how much of
the galaxy's ISM has already been stripped away, how rapidly the ISM
is being replenished by stellar mass loss within the galaxy, etc.  We
therefore content ourselves with using the X-ray morphology to measure
the direction in which each galaxy is moving on the plane of the sky.
In the remainder of this paper, we investigate whether such
information is sufficient to lift the degeneracy between orbital
anisotropy and gravitational potential.  In Section~2, we present a
practical measure of the orbital structure that can be determined
directly from the observed wake directions in a cluster.  Section~3
presents a simple dynamical model, which illustrates how the proposed
measure of orbital structure can discriminate between
otherwise-indistinguishable models.

\section{Estimating the projected anisotropy}

The direction in which the wakes in a cluster trail on the sky clearly
tells us something about the distribution of orbits.  Suppose, for
example, that the wakes imply that all the galaxies in the cluster are
moving either directly toward or away from the cluster centre in the
plane of the sky.  The only spherically-symmetric system that could
produce such an arrangement of projected motions is one in which the
galaxies are all on intrinsically radial orbits.  Thus, in this case
at least, the distribution of wake directions is sufficient to define
the distribution of orbits uniquely.  Note, however, that the relation
between the wake directions and orbital anisotropy is not completely
trivial: if all the galaxies follow circular orbits, the wakes will
{\em not}\/ all be tangentially-oriented on the plane of the sky,
and some will still appear pointed at the cluster centre.  

In order to explore the relationship between wake directions and
orbital distribution more fully, it is useful to define a ``projected
anisotropy.'' By analogy with equation~(\ref{eqn:beta}), an obvious
definition for this quantity is
\begin{equation}
B(R) = 1 - {\sigma_T^2 \over \sigma_R^2},
\label{eqn:Bprime}
\end{equation}
where $\sigma_T(R)$ and $\sigma_R(R)$ are the RMS velocities of
galaxies on the plane of the sky in the projected tangential and
radial directions, respectively.  

The projected anisotropy cannot be measured directly, but we can
estimate it from the observed distribution of projected orbit
directions.  If we define $\Theta$ to be the angle between the
direction of a galaxy's motion on the plane of the sky (as inferred
from its wake) and the projected radius vector of the galaxy, then the
quantity that we can observe directly is the distribution of projected
orbit directions as a function of projected radius, $P(\Theta, R)$.
This quantity can be related directly to $B(R)$.  For example, let us
assume that at some projected radius in the cluster, the plane-of-sky
velocity distribution (POSVD) can be approximated by a
function of the general form 
\begin{equation}
F(v_R,v_T) = F(v_R^2/a^2 + v_T^2/b^2),
\label{eqn:ellipsedist}
\end{equation}
where $a$ and $b$ are constants. By transforming from the coordinates
$\{v_R,v_T\}$ to polar coordinates on the plane of the sky, $\{v_{\rm
pos}, \Theta\}$, it is straightforward to show that the distribution
of observed values of $\Theta$ will be
\begin{equation}
P(\Theta) = {\sqrt{1-B} \over 2\pi(1 - B\cos^2\Theta)}.
\label{eqn:PTheta}
\end{equation}
Thus, we can estimate $B$ by fitting the functional form of equation
(\ref{eqn:PTheta}) to the observed distribution for $P(\Theta)$ at
any given projected radius.  

One robust method for implementing this fitting procedure is to
measure the fraction of galaxies for which $\pi/4 < \Theta < 3\pi/4$
or $5\pi/4 < \Theta < 7\pi/4$.  This quantity, $f_T$, measures the
fraction of galaxies whose orbits are closer to tangential than radial
on the plane of the sky.  By integrating equation~(\ref{eqn:PTheta}),
it is straightforward to show that the value of $B$ that gives rise to
a particular value of $f_T$ is
\begin{equation}
\hat B = 1 - \tan^2\left({\pi\over2} f_T\right).
\label{eqn:Bhat}
\end{equation}
Thus, by measuring $f_T$, we can use equation~(\ref{eqn:Bhat}) to
estimate the projected anisotropy.  This estimator for $B$ has a
number of desirable properties: it can be readily calculated from
observed wake directions; it is robust in the sense that a small
fraction of badly-determined directions will not affect it unduly; and
it has error properties which can be determined straightforwardly from
the binomial distribution of $f_T$.

Strictly speaking, $\hat B$ is only an unbiased estimator for $B$ if
the POSVD takes the elliptical form of
equation~(\ref{eqn:ellipsedist}).  However, a rather wide range of
plausible models can be approximated in this way (see below).  Even in
cases where this approximation is not valid, $\hat B$ still provides a
quantity that can be readily determined not only for observational
data, but also for dynamical models containing different orbit
distributions.  Thus, by comparing the observed profile $\hat B(R)$ to
that predicted by a dynamical model, it is possible to test the model
against observation.

\section{Case study: the Michie/truncated singular isothermal sphere 
model}

To see how useful $\hat B(R)$ might prove as a diagnostic of orbital
structure, we now consider a simple family of dynamical models.  Such
models are defined by the distribution function for the cluster, $f$,
which specifies the density of galaxies as a function of both velocity
and position (i.e. the phase density) -- see Binney \& Tremaine (1987)
for a thorough discussion of distribution functions.  By Jeans
theorem, for a spherical system in equilibrium, the distribution
function depends only on the energy, $E$, and angular momentum, $J$,
of the point in phase space under consideration.  A relatively simple
distribution function that might approximate the properties of a real
cluster of galaxies is the Michie model,
\begin{equation}
f(E,J) = \cases{{\rm const.} [e^{-E/\sigma_0^2} - 1]
                              e^{-J^2/(2\sigma_0^2r_a^2)} &$E < 0$\cr
                                  0                         & otherwise,\cr}
\label{eqn:MichieDF}
\end{equation}
where $\sigma_0$ and $r_a$ are constants (Michie 1963, Binney \&
Tremaine 1987).  This function produces a model where the distribution
of orbits is isotropic at small radii, but becomes radial at large
radii.  Thus, it mimics what we might expect in a cluster of galaxies:
relaxation effects will have randomized the orbits at small radii in
the cluster, but galaxies that travel to large radii will not have
relaxed, and so their orbits will still reflect the radial infall by
which the cluster formed.  The quantity $r_a$ defines the
characteristic radius at which the transition from isotropic to radial
orbits occurs.

The other dynamical quantity that we must specify for this model
cluster is the gravitational potential, $\Phi(r)$, by which the
galaxies are confined.  Since clusters of galaxies are dominated by
dark matter, and since we are interested in varying the orbital
anisotropy and gravitational potential independently, we do not make
the customary assumption that mass follows light.  Instead, we impose
an external gravitational potential, and vary its parameters to assess
the impact of changing the distribution of mass in the cluster.  The
simplest potential that enables us to carry out such an investigation
is the truncated singular isothermal sphere (TSIS), which is produced
by a mass density distribution
\begin{equation}
\rho(r) = \cases{\sigma_0^2/(2\pi G r^2) &$r < r_0$\cr
                      0                 &otherwise,\cr}
\label{eqn:rhoTSIS}
\end{equation}
where the constant $\sigma_0$ sets a characteristic velocity scale for
this potential.\footnote{There is no fundamental reason why this
characteristic velocity should be the same as the one in equation
(\ref{eqn:MichieDF}), but we adopt this value as the simplest choice.}
The TSIS density distribution produces a gravitational potential
\begin{equation}
\Phi(r) = \cases{
          2\sigma_0^2
            \left[\ln\left({r\over r_0}\right) 
                         + {r_0 \over r_{\rm max}} - 1\right]&$r < r_0$\cr
         -2\sigma_0^2r_0
            \left({1\over r} - {1\over r_{\rm max}}\right) &otherwise.\cr}
\label{eqn:PhiTSIS}
\end{equation}
Here, we have set the arbitrary constant offset in the potential such
that $\Phi(r > r_{\rm max}) > 0$.  Thus, from
equation~(\ref{eqn:MichieDF}), all the members of the cluster in this
model are confined to the region $r < r_{\rm max}$.

\begin{figure}
\epsfxsize\hsize
\epsffile{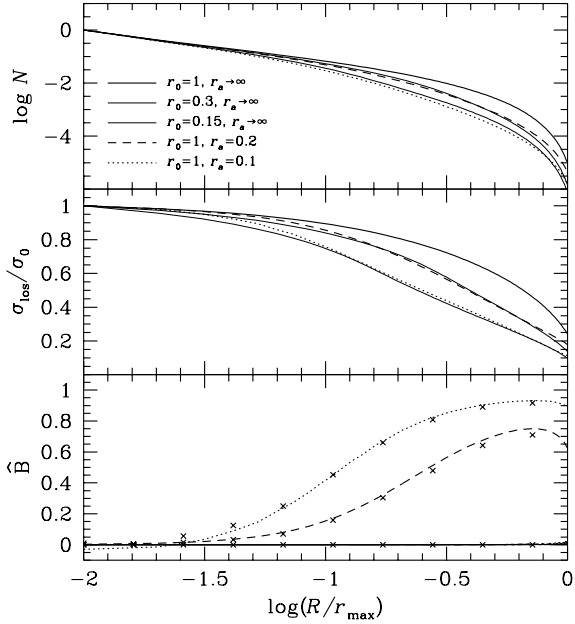}
\caption{\label{fig:modkin} The observable kinematics for the
dynamical models discussed in the text (with parameters $r_0$ and
$r_a$ as annotated).  Plotted as functions of projected radii, the
upper panel shows projected density, and the middle panel shows the
line-of-sight velocity dispersion.  The lower panel shows the
projected anisotropy: the lines show the estimator calculated using 
equation~(\protect\ref{eqn:Bhat}), while the crosses mark the
corresponding true projected anisotropy
[equation~(\protect\ref{eqn:Bprime})].  The central projected
densities and line-of-sight velocity dispersions have been scaled to a
common value.}
\end{figure}

Given $f$ and $\Phi$, we can calculate the density and velocity
distribution of galaxies at any point in the cluster.  It is thus
straightforward to calculate observable quantities -- such as $N(R)$,
$\sigma_{\rm los}(R)$, and the projected anisotropy estimator $\hat
B(R)$ -- by integrating the properties of the cluster along lines of
sight through the system at different projected radii $R$.
Figure~\ref{fig:modkin} shows the results for a set of such calculations
for the Michie distribution function in the TSIS potential.  The
spatial scale has been set by defining $r_{\rm max} = 1$, and various
combinations of the mass cut-off radius, $r_0$, and the orbital
anisotropy radius, $r_a$, have been adopted.

For the case $\{r_0 = 1, r_a \rightarrow \infty\}$, the mass cut-off
has no impact (since it lies outside the cluster), and so the
potential is that of a singular isothermal sphere.  At small radii,
the Michie distribution function is indistinguishable from an
isothermal distribution function, and so the cluster itself mimics a
singular isothermal sphere, with $N \propto R^{-1}$ and $\sigma_{\rm
los}$ approximately constant.  At larger radii, the more energetic
galaxies, which would have been included in an isothermal
distribution, are excluded from the Michie model, and so $N$ drops
faster than $R^{-1}$, and $\sigma_{\rm los}$ starts to fall.  Both
these quantities reach zero at $R = r_{\rm max}$, where the cluster
ends.

As $r_0$ is decreased, the truncation of the mass distribution affects
both $N$ and $\sigma_{\rm los}$.  The models with $r_0 = 0.3$ and $r_0
= 0.15$ have insufficient mass to contain the highest-velocity cluster
members, and so the density profile and the velocity dispersion both
drop more rapidly with radius.  Thus, both the dispersion profile and
density profile decline rapidly when the edge of the mass distribution
is reached.

However, as discussed in the introduction, the observed kinematics are
affected by orbital anisotropy as well as the shape of the potential.
As Fig.~\ref{fig:modkin} illustrates, when $r_a$ is reduced to a
finite value, the absence of galaxies on tangential orbits at large
radii means that the projected density drops more rapidly than for the
isotropic case.  Further, since galaxies on radial orbits travel
mostly transverse to the line-of-sight when seen at large projected
radii, the observed line-of-sight velocity dispersion is also
depressed.

In fact, as Fig.~\ref{fig:modkin} shows, it is possible to produce
very similar profiles for $N(R)$ and $\sigma_{\rm los}(R)$ using very
different combinations of $r_0$ and $r_a$.  In the case of the $\{r_0
= 1.0, r_a = 0.1\}$ and the $\{r_0 = 0.15, r_a \rightarrow \infty\}$
models, for example, the values of $N$ differ by less than $\sim 10\%$
inside $R = 0.8 r_{\rm max}$.  Within this radius, the values of
$\sigma_{\rm los}$ returned by the two models differ by less than
$5\%$.  Models with less extreme choices of these parameters, which
include both a cut-off in the density profile and some degree of
orbital anisotropy, differ even less.

\begin{figure}
\epsfxsize\hsize
\epsffile{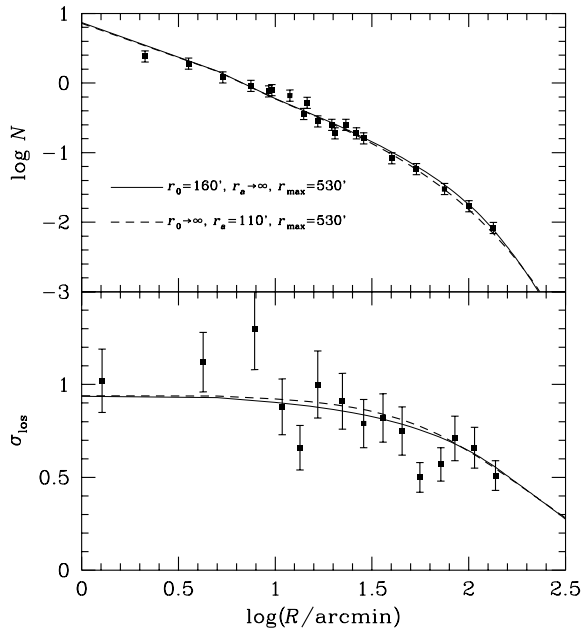}
\caption{\label{fig:coma} The projected density and line-of-sight 
velocity dispersion as a function of projected radius for Coma cluster
galaxies (Kent \& Gunn 1982) as compared to two of the models
discussed in the text.}
\end{figure}

The futility of trying to distinguish between such models is
illustrated in Fig.~\ref{fig:coma}.  This figure shows the kinematic
properties of the Coma cluster as derived by Kent \& Gunn (1982) using
line-of-sight velocities of more than 200 galaxies, with two of the
models from Fig.~\ref{fig:modkin} superimposed.  It should be stressed
that these models have not been formally fitted to the data, but
rather they are simply the arbitrarily-chosen models calculated above,
with their number-, velocity-, and length-scales chosen to match the
observations.  Indeed, it would appear that the models fail to
reproduce the observed flattening of the density profile at small
radii.  However, Beers \& Tonry (1986) have shown that such a flat
core is an artifact that arises from errors in the choice of cluster
centre.  With the centres measured correctly, they discovered that the
projected density profiles of most clusters contain no core at small
radii, but continue to rise in a power-law cusp, $N \propto R^{-1}$,
just as the models do.  What is clear from Fig.~\ref{fig:coma} is that
there is no way that the very different dynamical models can be
discriminated between on the basis of these data.  With very much
larger data sets, the error bars could be beaten down to a point where
the differences might become significant.  However, it is already
apparent from the lower panel of Fig.~\ref{fig:coma} that there are
likely to be systematic errors in the observed properties: modest
amounts of substructure in the cluster will mean that the simple
spherically-symmetric model is at some level invalid, and
contamination by unrelated galaxies along the line of sight may bias
the derived properties.  The presence of these fundamental systematic
uncertainties means that the collection of the vast data set necessary
to beat down the random errors would be a nugatory exercise.

Figure~\ref{fig:modkin} shows how the measurements of wake directions,
and hence $\hat B(R)$, might be used to lift this effective degeneracy
between the different dynamical models.  For the models in which $r_a
\rightarrow \infty$, the distribution of orbits is isotropic at every
point in the cluster.  Thus, in these cases there are as many galaxies
moving on radial orbits as tangential ones on the plane of the sky,
the distribution of wake directions is isotropic, and hence $\hat B
\equiv 0$.  However, for the models with finite values of $r_a$, the
radial orientation of the orbits is directly reflected by the positive
values of the projected anisotropy parameter.  Although the POSVD in
this model is not of the form specified by equation
(\ref{eqn:ellipsedist}), it is clear that $\hat B(R)$ is nonetheless a
good approximation to $B(R)$ for this moderately-realistic
distribution function; the two quantities never differ by more than
$\sim 10\%$.  Thus, the observable $\hat B(R)$ does provide a useful
measure of the underlying projected anisotropy.

The finite values of $\hat B$ predicted by the anisotropic models
would be readily detectable with relatively small samples of wake
directions.  If, for example, we assume that the cluster is isotropic,
then $f_T = 0.5$, and we can apply binomial statistics to show that a
sample of $N_{\rm wake}$ measured wake directions will produce an
estimate for this quantity of $\hat f_T = 0.5 \pm 0.5/\sqrt{N_{\rm
wake}}$.  The value of $\hat B \approx 0.7$ in the anisotropic models
shown in Figure~\ref{fig:modkin} corresponds to $f_T \approx 0.3$.
Thus, if the cluster were in reality better modeled by one of these
anisotropic systems, then a sample of only $N_{\rm wake} \approx 50$
wake directions drawn from projected radii where $\hat B \approx 0.7$
would be sufficient to rule out the isotropic $f_T = 0.5$ model at the
$3\sigma$ level.

\section{Discussion}

The passage of a galaxy through its host cluster must have some impact
on the distribution of the gas in its vicinity, and we are now finding
the first signs of these effects in X-ray observations.  The purpose
of this paper has been to address the question of whether the simplest
kinematic information that can be extracted from observations of this
phenomenon -- the distribution of directions of galaxy motion on the
plane of the sky -- can be useful in dynamical studies of clusters.
By means of the simple Michie/TSIS model, we have shown that such
observations can readily discriminate between models that have
different underlying orbit distributions and gravitational potentials,
but that are indistinguishable using traditional kinematic analyses.

In practice, the proposed measure of orbital anisotropy on the plane
of the sky, $\hat B(R)$, may not make optimal use of the observations,
since it requires that we bin the data radially.  It also discards any
information that might be gleaned from coupling between directions on
the plane of the sky and line-of-sight velocities: it is, for example,
possible that galaxies with the largest velocities might lie
preferentially on radial orbits, and so there might be a positive
correlation between line-of-sight velocity and how close a galaxy's
orbit is to radial on the plane of the sky.  It would be
straightforward to exploit this information by doing a full maximum
likelihood fit to the observed wake directions and line-of-sight
velocities of a sample of galaxies in a cluster.  The advantage of the
approach adopted here is that $\hat B$ can be calculated in a manner
that is robust against the effects of a few bad data points, while
such erroneous data could seriously compromise a maximum likelihood
analysis.  Further, the close tie between $\hat B(R)$ and the
intrinsic anisotropy parameter, $\beta(r)$, means that this
simple-to-measure quantity can readily be interpreted in terms of
the orbital properties of the cluster.

It is interesting to note that the signature of anisotropy in a
cluster is likely to be found at large radii (see
Fig.~\ref{fig:modkin}).  This radial dependence is fortunate, since it
will be easiest to detect a wake near the outskirts of the cluster,
where we are not viewing the interactions between the galaxy and the
ICM through a large column of unrelated X-ray emitting gas.
Similarly, this technique is particularly appropriate to poor systems,
where the column of ICM along the line of sight is relatively small.
Further, in poorer systems the energy density stored in the ICM is
lower, and so the disturbance caused by the passage of a galaxy will
be larger and hence more readily detectable.  It is also noteworthy
that contamination by non-cluster galaxies along the line of sight is
one of the major problems that dogs studies of low density regions
such as poor clusters and the outskirts of rich clusters.  Since such
unrelated galaxies will not interact with the ICM, it will be possible
to screen them out from the proposed studies based on their lack of
wakes.  The applicability of the technique to large radii in rich
clusters and to poor clusters is particularly exciting, since, as was
discussed in the introduction, it is in exactly these areas that the
existing methods for mapping the gravitational potential break down.

One possible difficulty in using wake directions to study cluster
dynamics is that the detectability of a wake may vary systematically
with the dynamics of the galaxy producing it.  For example, galaxies
traveling rapidly through the cluster will produce longer and more
dramatic wake phenomena, which will be easier to observe.  Similarly,
galaxies that have only recently joined the cluster are more likely to
contain a significant ISM, which will be readily detectable as a wake
of stripped material.  This possible bias can be minimized by
obtaining as deep an X-ray image as possible, so that even the fainter
wakes are detected.  Fortunately, an imaging X-ray telescope will
record many wakes simultaneously, and so the mapping process can be
carried out efficiently in few exposures.  Hopefully, with its high
spatial resolution and sensitivity, AXAF will open this new window on
the dynamics of clusters of galaxies.

\section*{ACKNOWLEDGEMENTS} 
 
The author gratefully acknowledges the support of a PPARC Advanced
Fellowship (B/94/AF/1840).

\end{document}